\documentclass[conference]{IEEEtran}
\IEEEoverridecommandlockouts
\usepackage{cite}
\usepackage{amsmath,amssymb,amsfonts}
\usepackage{algorithmic}
\usepackage{graphicx}
\usepackage{textcomp}
\usepackage{xcolor}
\usepackage{url}
\usepackage{hyperref}
\usepackage{pgfplots}

\usepackage{subfigure}
\usepackage{multirow}
\usepackage{caption,tabularx,booktabs}
\usepackage{booktabs}
\usepackage[export]{adjustbox}

\usepackage{comment}
\usepackage{placeins}

\def\BibTeX{{\rm B\kern-.05em{\sc i\kern-.025em b}\kern-.08em
		T\kern-.1667em\lower.7ex\hbox{E}\kern-.125emX}}
\begin{document}
	
	\title{SGX-MR-Prot: Efficient and Developer-Friendly Access-Pattern Protection in Trusted Execution Environments}
	
	\author{
		\IEEEauthorblockN{A K M Mubashwir Alam}
		\IEEEauthorblockA{\textit{Department of Computer Science} \\
			\textit{Marquette University}\\
			Milwaukee, WI \\
			mubashwir.alam@marquette.edu}
		\and
		\IEEEauthorblockN{Justin Boyce}
		\IEEEauthorblockA{\textit{Department of Computer Science} \\
			\textit{University of Arkansas}\\
			Fayetteville, AR \\
			jaboyce@uark.edu}
		\and
		\IEEEauthorblockN{Keke Chen}
		\IEEEauthorblockA{\textit{Department of Computer Science} \\
			\textit{Marquette University}\\
			Milwaukee, WI \\
			keke.chen@marquette.edu}
	}
	
	\maketitle
	
	\begin{abstract}
	Trusted Execution Environments, such as Intel SGX, use hardware supports to ensure the confidentiality and integrity of applications against a compromised cloud system. However, side channels like access patterns remain for adversaries to exploit and obtain sensitive information. Common approaches use oblivious programs or primitives, such as ORAM,  to make access patterns oblivious to input data, which are challenging to develop. This demonstration shows a prototype SGX-MR-Prot for efficiently protecting access patterns of SGX-based data-intensive applications and minimizing developers' efforts. SGX-MR-Prot uses the MapReduce framework to regulate application dataflows to reduce the cost of access-pattern protection and hide the data oblivious details from SGX developers. This demonstration will allow users to intuitively understand the unique contributions of the framework-based protection approach via interactive exploration and visualization.
	\end{abstract}
	
	\begin{IEEEkeywords}
	TEE, SGX, MapReduce, Data Analytics, Dataflow, Access Patterns, ORAM
	\end{IEEEkeywords}
	
	\section{Introduction}
	\label{sec:intro}
	With the development of resource-starving applications in big data, artificial intelligence, and the Internet of Things (IoT), public clouds have become a popular computing platform for their scalable storage and computation capacities. However, when uploading data and conducting computations in the cloud, data owners lose complete control of their data. They must fully trust service providers to take care of security, which raises significant concerns. To boost cloud users' trust and confidence, service providers have promoted the concept of \emph{confidential computing}, e.g., the trusted execution environment (TEE) in the cloud. TEEs aim to address the challenge of ensuring security for \emph{data in use}. Furthermore, due to their unique hardware-level mechanisms, TEEs promise to be resilient to attacks under extreme scenarios, i.e., adversaries fully controlling operating systems or hypervisors. Thanks to the hardware-assisted mechanism, confidential computing with TEEs is much more efficient than pure software cryptographic approaches.  
	
	Researchers are now focused on the threats from side channels, e.g., interactions with untrusted memory/storage areas, caches, and CPU micro-architectures \cite{nilsson20}. Attacks on caches and CPU micro-architecture are general side channels, not specific to TEEs, mostly depending on firmware-level patches to address. In contrast, the inevitable interactions between TEEs and untrusted memory/storage areas, i.e., \emph{access patterns}, are unique to the TEE approach and left to application developers to protect. Recent studies have shown significant risks due to leaked access patterns, especially for data-intensive applications \cite{olga16}. 
	
	Developers can manually compose oblivious solutions with oblivious primitives, such as oblivious branching and oblivious RAM (ORAM) \cite{olga16,sasy18}. However, it requires developers to understand the sensitive access patterns in their programs and comprehensively capture and handle them. Furthermore, developers must carefully select different oblivious primitives, e.g., ORAM-based sorting vs. oblivious sorting, to optimize the overall performance. All these factors make the manual composition approach time-consuming, error-prone, and unfriendly to normal developers. Automated approaches like Raccoon \cite{rane15} experimented with specialized compilers to convert regular programs. However, they are too complex to be practical yet. 
	
	Other work \cite{dinh15,zheng17} also tried to slightly modify existing data-intensive parallel processing software such as MapReduce and Spark to use TEEs and address identified sensitive access patterns. However, such top-down find-and-patch effects cannot systematically address the access pattern problem.
	
	\textbf{Contributions.} We have developed the SGX-MR approach \cite{alam21} to address the above problems. The basic idea is to use an application framework to regulate the application data flow. We can then identify the framework-level sensitive access patterns and develop methods to prevent the leakage of such patterns. We have shown that this approach (using MapReduce, for instance) can be more efficient than simple approaches like converting all I/O operations with ORAM \cite{sasy18}. It also simplifies users' TEE development by hiding most details of access-pattern protection, benefiting numerous data-intensive applications that can be cast to MapReduce. While we have used Intel Software Guard Extension (SGX) for our demonstration, the basic idea can be extended to address the TEE access pattern problem in general, e.g., for AMD SEV.

	This demo will use the prototype system SGX-MR-Prot to highlight the essential idea of SGX-MR and its working mechanism via a poster, interactive operations, and visualizations. It consists of the following major components: (1) the client-side user interface. It allows users to compile and deploy sample programs and observe the details of data encoding and program execution in the cloud, (2) the visualization of the cloud-side dataflow regulated by SGX-MR and the access patterns with and without SGX-MR protection, and (3) the performance comparison of SGX-MR with the ZeroTrace approach \cite{sasy18}, an efficient ORAM implementation for SGX. We believe this demo will promote the idea of framework-based access-pattern protection and possibly attract more researchers to join this effort.

	\begin{figure}
		\begin{minipage}{\linewidth}
			\centering
			\includegraphics[width =  0.7\linewidth]{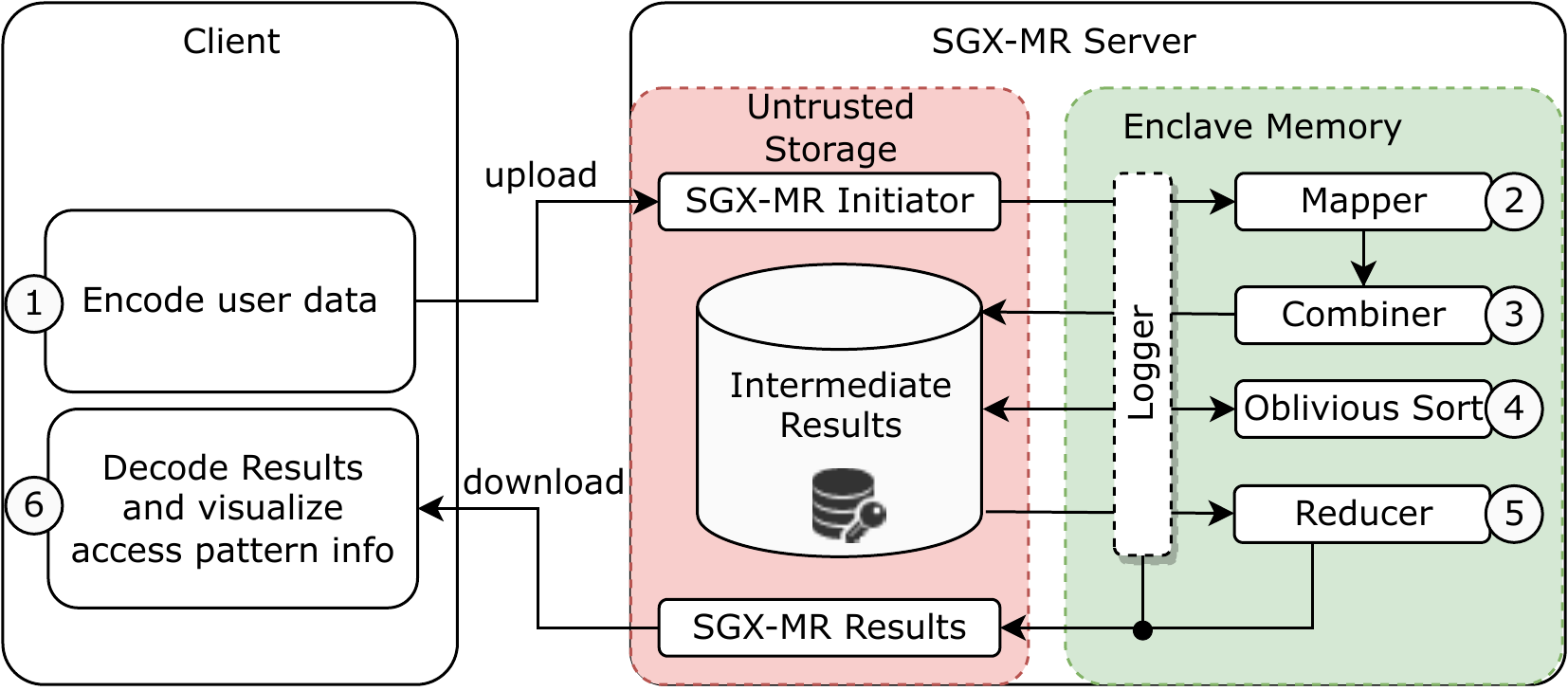}
			\caption{Workflow of the major components in the SGX-MR-Prot demo system}\label{fig:components}
		\end{minipage}
	\end{figure}
	
	\section{SGM-MR Approach}

	\subsection{Architecture}
	\textbf{Threat Model.} TEEs like SGX have a very small trusted computing base (TCB), which consists only of the CPU and the hardware-protected memory areas. A TEE program cannot be accessed directly by any other users, the OS, and the hypervisor. It assumes all hardware or software components other than the CPU and the protected memory can be malicious or compromised by the adversary. However, TEEs do not protect against side-channel attacks. While many side-channel attacks have to depend on firmware patches (e.g., the recent Spectra and Meltdown patches), access patterns between TEEs and non-TEE components have to be handled by developers.

	\textbf{Design Idea.} Our basic idea is to regulate the data flow of a class of data-intensive algorithms (i.e., machine learning algorithms) with an application framework to address the access pattern from the framework level that will benefit the whole course of applications. We also hide the protection details with the framework and minimize the developers' efforts in access pattern protection. Since MapReduce has been adopted for over a decade, we will use MapReduce to regulate application data flows.

	\textbf{SGX-MR Components.} 
	We assume that the SGX-MR server will be deployed in a public cloud that supports intel SGX, such as Microsoft Azure. According to the SGX working mechanism and features of data-intensive applications, we partition the entire framework into two parts, i.e., the trusted (in enclave) and untrusted parts. 
	
	\begin{itemize}
		\item \textbf{Untrusted Part}. The data and the I/O library are in the untrusted area. We design a block file structure for encrypted data on disk and in the untrusted memory area. The block size is tunable by the user based on the specific application but aligned to the page size, i.e., times of 4KB. Record length can be fixed or variable up to the user's setting. Each block contains a message authentication code (MAC) to ensure data integrity.
		
		\item \textbf{Enclave Part}. The remaining components of the SGX-MR framework reside in the enclave. The SGX-MR controller handles MapReduce jobs and controls the data flow of the application. Users only provide the map(), reduce(), and combine() functions to implement the application-specific processing. The prototype system will focus on the aggregation-style combine and reduce functions, such as COUNT, MAX, MIN, SUM, TOP-K, etc., to simplify the implementation, which has been shown sufficient to handle many data analytics tasks \cite{roy10}. The lightweight SGX-MR library can be compiled and linked with the application code (about 1.6 MB for sample programs).
		
	\end{itemize}

	\textbf{Application Workflow.}
	Figure \ref{fig:components} shows the data flow and interactions between system components. First, the input data file is processed by the data owner,  encoded with the block format via a \emph{file encoding} utility tool, and uploaded to the target machine, the application. Second, all file access requests from the enclave (i.e., mapper reading and reducer writing) must go through the SGX-MR block I/O module running in the untrusted memory area. Third, the intermediate outputs, e.g., of the Map, Combiner, and Sorting phases, can be spilled out in encrypted form by the application buffer manager or the system's virtual memory manager. The above data flow keeps the same for all applications that can be cast onto the MapReduce processing framework. Thus, we can effectively protect the access patterns for a broad category of data-intensive SGX applications by addressing the potential access-pattern leakages in the MapReduce dataflow.
	
	\subsection{Framework-Level Access Patterns and Protection }
	Based on the regulated application dataflow, we identify several critical data access patterns pertaining to different phases: Map's input, intermediate processing, combining, and output; Shuffling/Sorting's input, Sorting, and output; and Reduce's input, aggregation, and output. Among these access patterns, Map's input and Reduce's output involve only sequential block reads/writes. Thus, they do not leak additional information except for input and output file sizes (i.e., the number of blocks). We focus on other stages to identify sensitive access patterns and their mitigation methods.

	\subsubsection{Access Patterns in Sorting}
	The most expensive part of the whole data flow is the Sorting phase. Traditionally, the MapReduce framework adopts the MergeSort algorithm for its simplicity. As the shares of Map(or combiner)-output has been sorted individually, MergeSort only needs to merge the sorted shares. However, we have identified some unusual merging behaviors, which reveal sensitive information about record groups -- the ordering and the size of the group. For example, in the WordCount program, adversaries might be able to guess the frequencies of words and derive the word distribution. 
	
	\textbf{Mitigation Methods.} This access pattern problem involves two parts: the block-level pattern in untrusted memory and the page-level pattern in in-enclave memory. We address the block-level patterns with \emph{oblivious Sorting} algorithms. The in-enclave page-level patterns will be discussed later. An oblivious sorting algorithm takes the fixed identical access pattern that is only determined by the size of the list invariant to input data values. We implement the well-known \emph{BitonicSort}  algorithm for our framework.

	\subsubsection{In-Enclave Page Access Patterns}
	Enclave execution is vulnerable to the page-fault attack, as discovered by previous studies \cite{shinde16}. Page-fault attack aims at branching statements, in particular, the in-enclave sorting part of the Sorting phase and the map-output Sorting (when a block is filled up) before combining. The repetitive page accesses for the same data block can reveal the ordering of records in a pair of blocks.
	
	\textbf{Mitigation Methods.} The core operation of the in-enclave BitonicSort, \emph{compare-and-swap}, can be captured by observing the sequence of page faults and used to reveal the relative ordering between pairs of records. The following code snippet shows how the access pattern is possibly associated with the record order. 
	
	\begin{verbatim}
		if (a >= b){
			// swap a and b, and 
			// page access is observed.
		}else{
			// no page access.}
	\end{verbatim}
	
	To avoid this access-pattern leakage, we adopt the oblivious swap operation \cite{olga16}, which uses the CMOV instructions to hide the page access patterns.
	
	\subsubsection{Leakage in Reducing} 
	\label{sec:leakage-in-reducer}
	The Sorting phase orders key-value pairs by the key and sorts them into groups. The controller sequentially reads the sorted blocks and transfers the records containing the same key (i.e., a group of records) to the reduce function, which will reduce the group to one key-value pair. By observing the input/output ratio, adversaries can estimate group sizes.

	\textbf{Mitigation Methods.} In the demo system, we have made the Combiner component mandatory for the aggregation-based reduce functions. With Combiners, each record entering the Reduce phase may correspond to the pre-aggregate of multiple forms. Thus, counting the records in the Reducer input does not estimate actual group sizes well. Dummy records (with count zero) are also randomly injected into the Combiner result to further disguise group sizes.
	
	\section{Demonstration}
	\label{sec:demo}
	The whole demonstration consists of the following components:
	
	\textbf{Introduction.} The first part of the demonstration uses a poster and slides to outline the unique features of SGX-MR and introduces the viewer to the SGX-MR-Prot demonstration system. 
	
	\textbf{Live System.} The user will be able to use the client-side system to encode the raw data, select a sample analytics algorithm, such as KMeans or WordCount, submit a request to the server, and observe how the algorithm is securely processed by the server with the help of visualizations and statistics. 
	
	\textbf{Comparison with Existing Approaches.} We will also compare the performance with the ZeroTrace approach \cite{sasy18} that is based on an efficient implementation of ORAM. This helps users understand the efficiency and unique advantages of the proposed approach.
	
	Next, we give more details about the workflow, the live system, and the approach comparison. 
	
	\begin{figure}[t]
		\centerline{\includegraphics[width=0.7\linewidth]{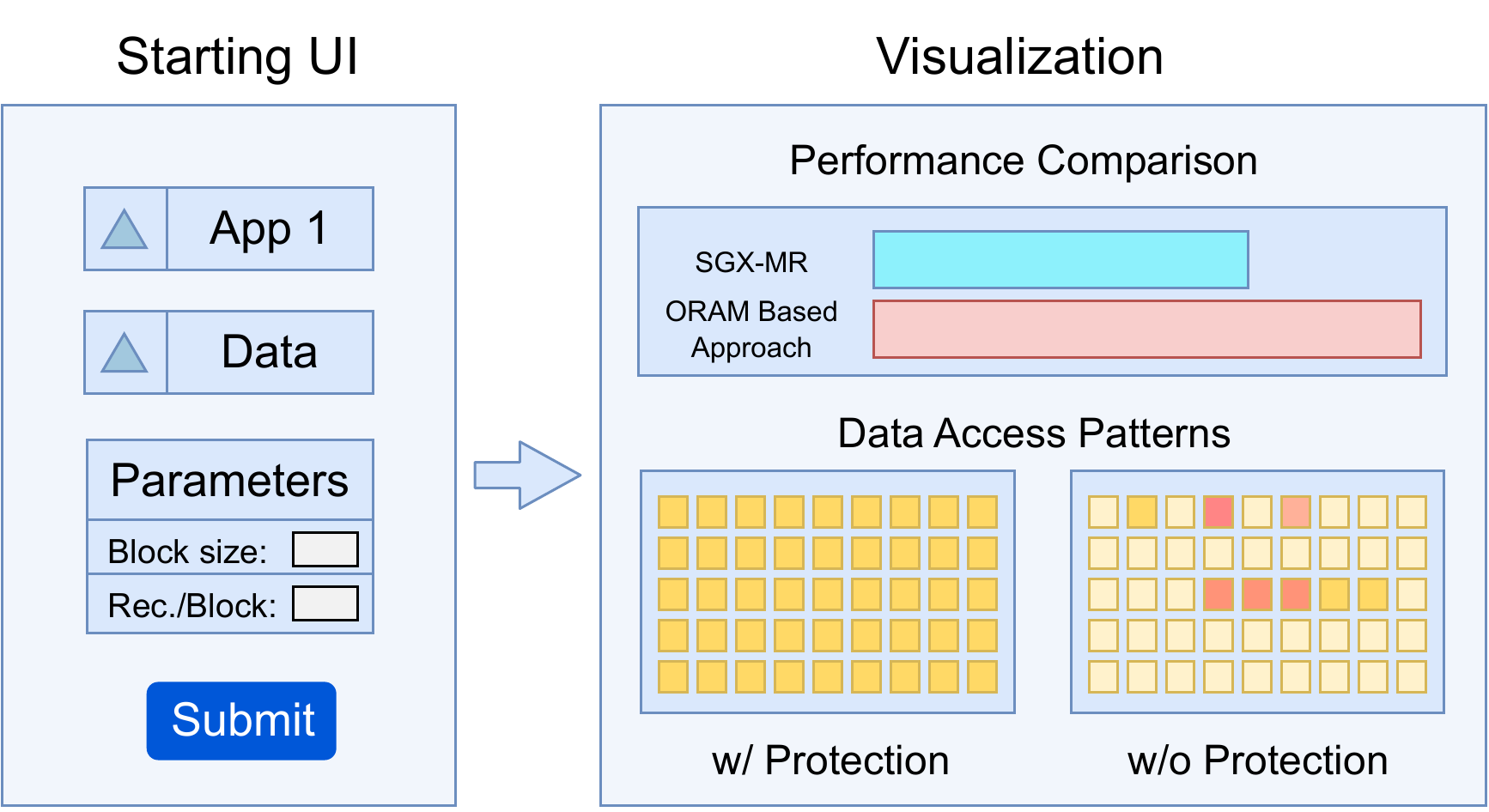}}
		\caption{Sample design of the interactive demo and visualization components.}
		\label{fig:ui}
	\end{figure}

	\subsection{Demonstration Workflow}
	\begin{enumerate}
		\item After selecting the application and dataset, the client application sends the encrypted data blocks to the server. Server then stores the data in untrusted memory first and then sequentially transfers the blocks to the enclave program, where the SGX-MR controller runs the selected application, generates the encrypted result, and saves it back to the untrusted area. Finally, a data transfer tool in the untrusted area delivers the result to the client.
		
		\item When the cloud side is processing data, the demo system will log the intermediate processing information and send it back to the client-side demo system. The client demo system uses this information to visualize the processing steps and the access patterns with or without the protection mechanism provided by SGX-MR.
		
		\item  We also developed the sample applications using the ZeroTrace open-source implementation \footnote{https://github.com/sshsshy/ZeroTrace}. The cloud-side demo system also runs the ZeroTrace-based applications to generate results for comparison. The demo system keeps track of the amount of memory and time used by our and the ZeroTrace approaches and shows the comparison to the user.  
		
		\item We provide a few sample applications, such as WordCount, kMeans, and PageRank, in the demo. We will show how easily application developers can program such applications without any knowledge of TEE and access patterns. 
	\end{enumerate}
	
	\subsection{Implemented Functionality} 
	The demo system implements the major components in Figure \ref{fig:components}, including data encoding, decoding, key-information logging, and visualizations of the intermediate processing steps and access patterns. The server processing components have been implemented with C++ and deployed in an SGX-enabled system, while the client interface is implemented with JavaScript. 
	
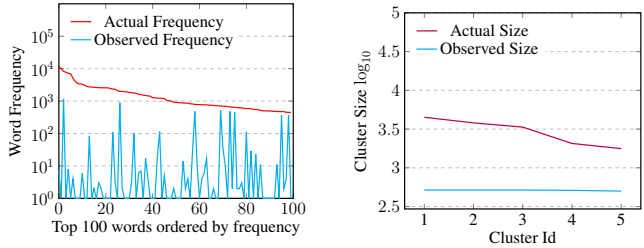
\begin{figure}[h]
		\centering
		\begin{tabular}{p{0.47\linewidth}p{.45\linewidth}}
			
			\subfigure[Word frequencies in WordCount problem]{
				\resizebox{\linewidth}{!} {
					\begin{tikzpicture}
						\begin{axis}[
							line width=1pt,
							scaled ticks=false,
							xlabel={Top 100 words ordered by frequency},
							ylabel={Word Frequency},
							xmin=0, xmax=100,
							ymin=1, ymax=1000000,
							xtick={0, 20, 40, 60, 80, 100},
							ytick={1, 10, 100, 1000, 10000, 100000},
							label style={font=\Large},
							tick label style={font=\Large},
							legend style={ draw = none, font=\Large},
							legend pos=north west,
							ymajorgrids=true,
							grid style=dashed,
							ymode=log,
							log basis y={10}
							]
							
							\addplot[
							color=red,
							]
							coordinates {
								(0, 11841)
								(1, 9809)
								(2, 8291)
								(3, 7681)
								(4, 7198)
								(5, 6956)
								(6, 5069)
								(7, 4023)
								(8, 3493)
								(9, 3361)
								(10, 3334)
								(11, 3067)
								(12, 2821)
								(13, 2743)
								(14, 2688)
								(15, 2669)
								(16, 2629)
								(17, 2594)
								(18, 2577)
								(19, 2576)
								(20, 2560)
								(21, 2557)
								(22, 2425)
								(23, 2343)
								(24, 2322)
								(25, 2085)
								(26, 1994)
								(27, 1992)
								(28, 1938)
								(29, 1922)
								(30, 1849)
								(31, 1834)
								(32, 1762)
								(33, 1744)
								(34, 1622)
								(35, 1589)
								(36, 1524)
								(37, 1487)
								(38, 1447)
								(39, 1415)
								(40, 1264)
								(41, 1258)
								(42, 1233)
								(43, 1226)
								(44, 1213)
								(45, 1208)
								(46, 1041)
								(47, 1036)
								(48, 988)
								(49, 921)
								(50, 921)
								(51, 897)
								(52, 883)
								(53, 875)
								(54, 869)
								(55, 864)
								(56, 829)
								(57, 814)
								(58, 790)
								(59, 782)
								(60, 778)
								(61, 771)
								(62, 769)
								(63, 764)
								(64, 747)
								(65, 744)
								(66, 737)
								(67, 719)
								(68, 714)
								(69, 708)
								(70, 693)
								(71, 689)
								(72, 679)
								(73, 658)
								(74, 650)
								(75, 642)
								(76, 628)
								(77, 625)
								(78, 614)
								(79, 597)
								(80, 590)
								(81, 588)
								(82, 584)
								(83, 553)
								(84, 552)
								(85, 538)
								(86, 522)
								(87, 500)
								(88, 499)
								(89, 498)
								(90, 496)
								(91, 488)
								(92, 481)
								(93, 481)
								(94, 476)
								(95, 473)
								(96, 473)
								(97, 453)
								(98, 449)
								(99, 446)
							};
							\addlegendentry{Actual Frequency}
							
							\addplot[
							color=cyan,
							]
							coordinates {
								(0, 7)
								(1, 1)
								(2, 1168)
								(3, 1)
								(4, 8)
								(5, 1)
								(6, 4)
								(7, 1)
								(8, 1)
								(9, 1)
								(10, 6)
								(11, 1)
								(12, 1)
								(13, 85)
								(14, 1)
								(15, 2)
								(16, 1)
								(17, 3)
								(18, 2)
								(19, 1)
								(20, 1)
								(21, 1)
								(22, 1)
								(23, 109)
								(24, 3)
								(25, 1)
								(26, 892)
								(27, 2)
								(28, 1)
								(29, 1)
								(30, 1)
								(31, 3)
								(32, 100)
								(33, 1)
								(34, 6)
								(35, 7)
								(36, 1)
								(37, 17)
								(38, 4)
								(39, 1)
								(40, 1)
								(41, 1)
								(42, 19)
								(43, 114)
								(44, 1)
								(45, 5)
								(46, 1)
								(47, 1)
								(48, 1)
								(49, 2)
								(50, 1)
								(51, 1)
								(52, 1)
								(53, 14)
								(54, 2)
								(55, 4)
								(56, 1)
								(57, 26)
								(58, 480)
								(59, 10)
								(60, 1)
								(61, 4)
								(62, 6)
								(63, 17)
								(64, 1)
								(65, 1)
								(66, 2)
								(67, 1)
								(68, 1)
								(69, 516)
								(70, 35)
								(71, 10)
								(72, 1)
								(73, 481)
								(74, 2)
								(75, 446)
								(76, 2)
								(77, 2)
								(78, 3)
								(79, 1)
								(80, 112)
								(81, 1)
								(82, 29)
								(83, 1)
								(84, 23)
								(85, 1)
								(86, 11)
								(87, 1)
								(88, 1)
								(89, 1)
								(90, 1)
								(91, 1)
								(92, 1)
								(93, 11)
								(94, 1)
								(95, 371)
								(96, 2)
								(97, 4)
								(98, 373)
								(99, 1)
							};
							\addlegendentry{Observed Frequency}
							
						\end{axis}
					\end{tikzpicture}
					\label{fig:word-freq}
				}
			}&
			
			\subfigure[Cluster sizes in KMeans algorithm.]{
				\resizebox{\linewidth}{!} 
				{
					
					\begin{tikzpicture}
						
						\begin{axis}[
							line width=1pt,
							scaled ticks=false,
							xticklabels={1, 2, 3, 4, 5},
							ymax=5,
							ylabel style={align=center},
							ylabel={Cluster Size $\log_{10}$},
							xlabel={Cluster Id},
							label style={font=\Large},
							tick label style={font=\Large},
							legend style={ draw = none, font=\Large},
							legend pos= north west,
							ymajorgrids=true,
							grid style=dashed,
							xtick=data,
							]
							
							\addlegendentry{Actual Size}            
							\addlegendentry{Observed Size}            
							\addplot [draw=purple]  coordinates {(1, 3.6521496054) (2,3.5798978696) (3, 3.52543355343) (4,3.31428866095) (5,3.24748226068)};
							\addplot [draw=cyan] coordinates {(1, 2.71264970163) (2, 2.71264970163) (3, 2.71264970163) (4,2.70926996098) (5, 2.69897000434) };
						\end{axis}
					\end{tikzpicture}
					\label{fig:cluster-size}
				}
			}
		\end{tabular}
		\caption{Exemplary comparison of group size results between what is actual and what adversaries can learn from SGX-MR access patterns.}
		\label{fig:group-size}
	\end{figure}

	A highlight of the demo system is the visualization components for users to understand the ideas behind the method intuitively. Figure \ref{fig:ui} shows a conceptual visualization of the demo client. With the server-side logging component, we can trace the system execution and show the observed access patterns between the enclave and the untrusted area. By turning on or off the access-pattern protection mechanism in SGX-MR, users can observe the change in access patterns and associated costs for protection. Figure \ref{fig:group-size} shows the application-level access pattern protection from SGX-MR's experimental result.

	\subsection{Performance Comparison}
	To further understand the advantages of the SGX-MR approach, we also demonstrate the comparison with the ORAM-based approach. We will utilize ZeroTrace, an open-source efficient ORAM implementation for SGX, for performance comparison. We will show the basic procedure of ZeroTrace-based application development and then run the sample application implemented with SGX-MR vs. that with ZeroTrace to observe the cost difference. Furthermore, we will also show the total memory needed by both methods. Our preliminary results \cite{alam21} have shown that our method has a significantly lower storage cost and processing time compared to ZeroTrace.
	
	\subsection{User Interactions}
	Interactivity is the key feature of this demo. We briefly describe a sample use scenario from the developer's perspective. 
	
	\textbf{Development.}
	The SGX-MR hides access-pattern protection details with the MapReduce processing layer. To develop a data-oblivious application, the developer only needs to write the map() and reduce() functions in C++ for the desired application and handle very few simple access patterns. We show sample code for simple applications like KMeans and WordCount. 
	
	\textbf{Compilation.} The demo system includes the framework code in a library. By executing a building script, the developer links the user-defined functions with the framework and generates a signed TEE binary. 
	
	\textbf{Deployment.} The developer uses a client-side encoder tool to convert a plaintext data file to an encrypted block file with specified block and record sizes. The client-side tool uploads the encoded data and the signed binary to the cloud. During and after the execution, the server demo program will send logged information to the client demo system.

	\section{Summary}
	We present the SGX-MR-Prot demo system to show that using an application framework to regulate dataflows and hide details from developers can be an efficient approach for TEE programs' access pattern protection. It will be highly interactive and visual, allowing users to understand the technical details easily and appreciate the advantages of this approach. 
	
	\section*{Acknowledgment}
	
	\small{ This research was partially supported by National Institute of Health (1R43AI136357) and National Science Fundation (2232824). Any opinions, findings, and conclusions or recommendations expressed in this material are those of the authors and do not necessarily reflect the views of the funders.}
	
	\bibliographystyle{abbrv}
	\bibliography{./paper,./sgx_sc}
	
\end{document}